\documentclass[mathleft]{an}
\usepackage{graphicx}
\usepackage{times}
\usepackage{caption}
\usepackage{amsmath}
\usepackage{wasysym}
\begin{document}

\Pagespan{1}{}
\Yearpublication{2017}
\Yearsubmission{2017}
\Month{12}
\Volume{1}
\Issue{1}
\DOI{10.1002/asna.201713365}

\title{Radial velocity measurements and orbit determination of eight single-lined spectroscopic binary systems\thanks{Based on observations obtained with telescopes of the University Observatory Jena, which is operated by the Astrophysical Institute of the Friedrich-Schiller University.}}

\author{R.~Bischoff\fnmsep\thanks{Corresponding author: \email{richard.bischoff@uni-jena.de}\newline}, M. Mugrauer, T. Zehe, D. W\"{o}ckel, A. Pannicke, O. Lux, D. Wagner, T. Heyne, C. Adam and  R. Neuh\"{a}user}

\titlerunning{RV measurements and orbit determination}
\authorrunning{Bischoff et al.}
\institute{Astrophysikalisches Institut und Universit\"{a}ts-Sternwarte Jena, Schillerg\"{a}{\ss}chen 2-3, 07745 Jena, Germany}

\received{24 Feb 2017}
\accepted{22 May 2017}


\abstract{Since early 2015 a new radial velocity monitoring campaign is going on at the University Observatory Jena. The aim of this project is to obtain current radial velocity measurements of selected single-lined spectroscopic binary systems, to redetermine and/or constrain their orbital solutions. In this paper we characterize the properties of the target sample of the project, describe the spectroscopic observations, the data-reduction and analysis and present the first results of the project, taken with the fiber-linked astronomical spectrograph FLECHAS. We present 391 radial velocity measurements of eight spectroscopic binaries, which were taken within an epoch difference of  1.6\,years. These radial velocities were used to determine the spectroscopic orbital elements of the observed binary systems, which exhibit orbital periods in the range between nearly one up to several dozens of days. We could constrain the orbital solutions of seven of these binary systems, and redetermine the orbital solution of $\varphi$\,Dra, whose orbit exhibits $>4$ times longer orbital period and is more eccentric than given in the 9th Catalogue of Spectroscopic Binary Orbits.}

\maketitle

\maketitle

\section{Introduction}

The University Observatory Jena is located close to the small village Gro{\ss}schwabhausen (GSH), west of the city of Jena (Pfau 1984). \\
The observatory is equipped with a 90\,cm-reflector that can be operated as a Nasmyth-telescope ($\text{D}=90$\,cm, $\text{f/D}=15$) for spectroscopic observations. In 2013 a new fibre-linked \'Echelle spectrograph (\cite{mugrauer2014}) was installed. Among the first observing campaigns carried out with FLECHAS (\cite{irrgang2016}), a new radial velocity (RV) monitoring program of selected spectroscopic binaries was started in early 2015 in addition to the Gro{\ss}schwabhausen binary survey (\cite{mugrauer2016}). The aim of this new project is to obtain current RV data of spectroscopic binaries and to redetermine or better constrain their orbital solutions. The binary systems selected for the RV monitoring campaign were taken from the \textsl{9th Catalogue of Spectroscopic Binary Orbits} (SB9 here after, see \cite{pourbaix2004}), which contains orbital elements of several thousand spectroscopic binary systems in its current version\footnote{The catalog is available online in the VizieR database, or at: \newline \texttt{http://sb9.astro.ulb.ac.be/mainform.cgi}}. The binaries in this catalog are listed with different orbit grades, which define the precision of their orbital solution, ranging from~0 for a poor to~5 for a definitive orbit determination. For the FLECHAS project we selected single-lined spectroscopic binaries from this database as targets, which (a) are observable throughout the whole year at airmasses $X<2.5$ ($\text{Dec}>+62^{\circ}$), (b) are bright enough ($V\leq6$\,mag) to obtain spectra with a sufficiently high signal-to-noise ratio $\text{(SNR)}>100$ with FLECHAS at short integration times of only a few minutes, (c) exhibit orbit grades up to~4, that are binary systems, whose orbital solutions still need to be correctly determined or be further constrained. In total, 20 targets were identified in the SB9 for the project and observations were carried out so far for those eight binary systems which exhibit the largest RV semi-amplitudes. The properties of the observed targets are summarized in Tab.\,\ref{table_targetproperties}.

In section 2 we will describe the observations carried out with FLECHAS. In the following section we present the obtained RV measurements of all observed spectroscopic binaries used to derive the orbital elements of these systems. The obtained orbital solutions of the systems are presented in section 4 and discussed in detail in the final section of this paper.

\begin{table*}
\centering \caption{The properties of the observed spectroscopic binaries. Their right ascension (RA), declination (Dec), visual magnitude (V), orbit grade (G), spectral type (SpT) and the orbital elements of the systems, namely their orbital period (P), RV semi-amplitude (K), systemic velocity ($\gamma$), orbital eccentricity (e), mass function $f(m)$ and semi-major axis $a\sin(i)$ with their reference (ref) as given in the SB9, are listed in the bottom table.}
\begin{tabular}{ccccccccccccc}
	\hline
	Target          & RA               & Dec\,            & V 		 & G	&  SpT     	&$P$        & $K$          & $\gamma$   & $e$  & $f(m)$			 & $a\sin(i)$ & ref\\
	
	& $[$hh mm ss.s$]$&$[$dd mm ss$]$&&&&$[$d$]$&$[$km/s$]$&$[$km/s$]$&&$[\text{M}_{\odot}]$&$[\text{au}]$&\\
	\hline
	CO\,Cam			& 12 12 11.9       & +77 36 58      & 5.1   & 3 &   A5m       	& 1.271   & 63.2       & +0.3       & 0    & 0.033 & 0.0074	&[a]\\
	HR\,791			& 02 44 49.7       & +67 49 29      & 6.0	& 3 &   A5III     	& 2.5364  & 55.1       & +4.3       & 0    & 0.044 & 0.0128	&[b]\\
	HR\,1401		& 04 33 30.7       & +72 31 43      & 5.9	& 3 &   A8m       	& 4.195   & 31.3       & +9.0	    & 0    & 0.013 & 0.0121	&[c]\\
	71\,Dra			& 20 19 36.7       & +62 15 27      & 5.6	& 2 &   B9V   	    & 5.29    & 49.7       & -7.8	    & 0.04 & 0.067 & 0.0242	&[d]\\
	$\alpha$\,Dra   & 14 04 23.3       & +64 22 33      & 3.7	& 4 &   A0III       & 51.4167 & 49.7       & -14.0      & 0.4  & 0.505 & 0.2153 &[e]\\
	$\omega$\,Cas	& 01 56 00.0       & +68 41 07      & 5.0   & 3 &   B8III       & 69.92   & 29.6       & -24.8      & 0.3  & 0.164 & 0.1815	&[f]\\
	OS\,UMa			& 08 19 17.2       & +62 30 26      & 5.7	& 3 &   G8III       & 89.0653 & 22.7       & -3.3	    & 0.19 & 0.103 & 0.1825	&[g]\\
	$\varphi$\,Dra  & 18 20 45.4       & +71 20 16      & 4.2	& 1 &   A0p         & 26.768  & 26.6       & -20.8      & 0.39 & 0.041 & 0.0603	&[h]\\
	\hline\\
\end{tabular}
\flushleft{
	\footnotesize{[a]\hspace{0.25cm}}\cite{lee1916}\hspace{1.5cm}
	\footnotesize{[b]\hspace{0.25cm}}\cite{luyten1936}\hspace{1.5cm}
	\footnotesize{[c]\hspace{0.25cm}}\cite{luyten1936}\hspace{1.5cm}
	\footnotesize{[d]\hspace{0.25cm}}\cite{hube1973}\\
	\footnotesize{[e]\hspace{0.25cm}}\cite{elst1983}~\hspace{0.35cm}
	\footnotesize{[f]\hspace{0.25cm}}\cite{young1915}~~~\hspace{1.35cm}
	\footnotesize{[g]\hspace{0.25cm}}\cite{carquillat1983}~~\hspace{0.4cm}
	\footnotesize{[h]\hspace{0.25cm}}\cite{abt1973}}
\label{table_targetproperties}                                              		
\end{table*}

\section{Observations and data reduction}	

In the course of our RV monitoring campaign between February 2015 and September 2016, spectroscopic data of all observed targets were taken with FLECHAS at multiple observing epochs (more than 40 for each target). For the spectroscopic observations the full resolving power of the spectrograph of $R\sim9300$ was used, and spectra with signal-to-noise ratio (SNR) between 120 and 385 (as measured at 6520\,\AA, dependent on the weather and atmospheric seeing conditions) were recorded for all observed spectroscopic binary systems, which was sufficiently high to obtain accurate RV measurements for all our targets. Further details of the observations are summarized in the observing log Tab.\,2.

In all observing epochs, three spectra of each target with an individual integration time of 150\,s were recorded, in order to achieve sufficiently high SNR and to optimally remove cosmics detected in the individual spectra. For calibration purposes, we have always taken three spectra of a ThAr-lamp (with more than 700 lines detected) for the wavelength calibration before the spectroscopy of each target. Furthermore, we obtained three well-exposed flat-field frames of a tungsten lamp each with an integration time of 5\,s and then started the observation of spectroscopic binaries. In addition, each night, dark frames with the corresponding integration times were taken in the evening or morning twilight.

The reduction of all spectroscopic data was done with the FLECHAS software pipeline, which was developed at the Astrophysical Institute Jena and is optimized for the reduction of FLECHAS data. The pipeline includes the dark-subtraction and flat-fielding of all spectroscopic data, as well as the extraction, wavelength calibration and averaging (including cosmic removal) of all individual spectral orders (\cite{mugrauer2014}).

\begin{table}
\centering\caption{Observation log. For each target we list the number of observing epochs ($\text{N}_{\text{Obs}}$), the dates of the first and last successful observation, as well as the average signal-to-noise ratio (ave. SNR) achieved in the spectroscopic observations, measured at 6520\,\AA.}
\begin{tabular}{ccccc}
\hline
Target          & $\text{N}_{\text{Obs}}$ & first epoch & last epoch  & ave. SNR \\
\hline
CO\,Cam			& 45                      & 2016 Feb 12 & 2016 Sep 27 & 225        \\
HR\,791			& 52                      & 2016 Feb 12 & 2016 Sep 27 & 120        \\
HR\,1401		& 50                      & 2016 Feb 12 & 2016 Sep 27 & 130        \\
71\,Dra			& 51                      & 2016 Feb 12 & 2016 Sep 27 & 150        \\
$\alpha$\,Dra   & 41                      & 2016 Feb 02 & 2016 Sep 27 & 385        \\
$\omega$\,Cas	& 43                      & 2016 Feb 12 & 2016 Sep 27 & 180        \\
OS\,UMa			& 40                      & 2016 Feb 12 & 2016 Sep 27 & 170        \\
$\varphi$\,Dra  & 69                      & 2015 Feb 22 & 2016 Aug 18 & 305        \\
\hline \\
\end{tabular}
\label{table_log}                                              		
\end{table}

So far, 391 RV measurements have been taken of the observed eight spectroscopic binaries, which results in a total on source integration time of about 49\,hr.

\section{Radial velocity measurements}

The Standard IRAF script \texttt{splot} for the analysis of spectroscopic data was used to determine the RVs of all observed targets at the individual observing epochs. The routine includes Gaussian fitting, which was used within the cores of the spectral lines, which are dominated by Doppler broadening, to determine the wavelength of the line centers.
In order to obtain the RVs of the targets, we determined the central wavelengths $\lambda$ of the hydrogen Balmer lines  ($H_{\alpha}: \lambda_{0} = 6562.81$\AA, $H_{\beta}: \lambda_{0} = 4861.34$\AA~and $H_{\gamma}: \lambda_{0} = 4340.48$\AA). The Balmer lines of all spectra are within the spectral range of FLECHAS and therefore well detected.

The measured shift of wavelength $(\lambda - \lambda_{0})$ of a spectral line yields the RV of the star:
\begin{equation}
RV = c \cdot\frac{\lambda - \lambda_{0}}{\lambda_{0}} + BC,
\end{equation}
where is $c$ the speed of light and $BC$ the barycentric correction which accounts for the RV offset between the observing site and the barycenter of the solar system. Consequently, each Date of Observation was corrected from the Julian Date to receive the Barycentric Julian Date (BJD).

The determined average RVs of all targets are illustrated in Fig.\,\ref{fig:plot} and are summarized in the Appendix in Tab.\,\ref{rv_cocam} to \ref{rv_phidra} for all observing epochs. The RV precision achieved in our project is in the level of 1\,km/s, on average, and ranges between 0.7 and 2.1\,km/s for the individual targets.

Besides the RV measurements, we also tested the stability of the wavelength calibration of the spectrograph by measuring the wavelengths of several selected telluric lines in all available spectra. No trends or strong deviations in the calibration were found throughout the observations, and the wavelength scatter measured for the telluric lines (0.7\,km/s to 1.8\,km/s) corresponded to an RV variation that is consistent with or even smaller than the achieved typical RV precision of the observed spectroscopic binary systems. Hence, the wavelength calibration of FLECHAS can be considered as stable in the long term.

\section{Orbit determination}

\begin{figure*}
	\centering\includegraphics[width=23cm,height=23cm,keepaspectratio]{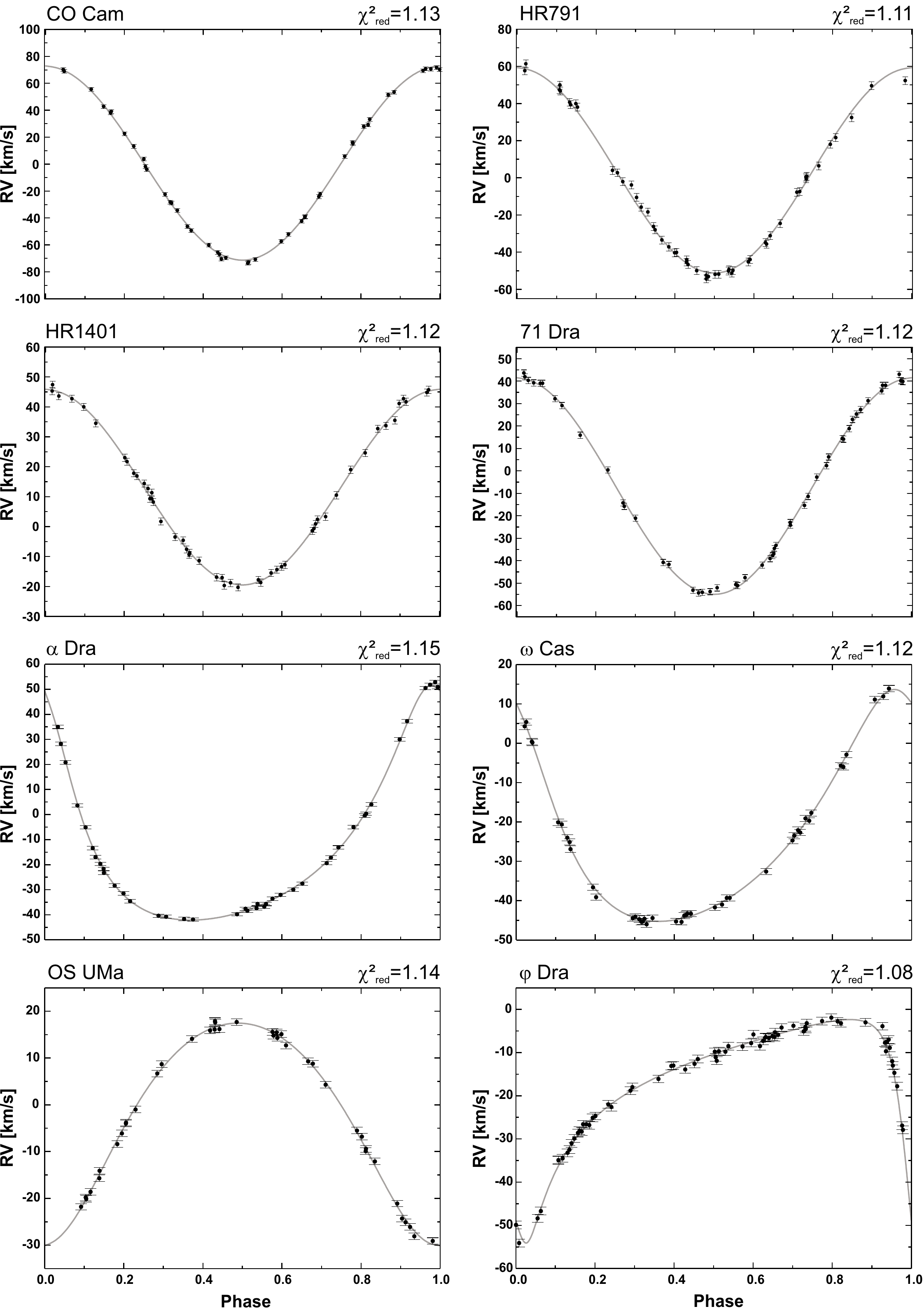}
	\caption{Phase folded RV curves and determined spectroscopic orbits (grey curved lines) of all observed targets.}
	\label{fig:plot}
\end{figure*}

The RV of the primary component on a spectroscopic orbit around the barycenter of a binary system can be calculated via
\begin{equation}
RV = K \cdot [e\cdot \cos (\omega) +\cos (\nu+\omega)]+\gamma,
\end{equation}
with the semi-amplitude $K$, the argument of periastron $\omega$, the eccentricity $e$, and the true anomaly $\nu$ of the star on its orbit, as well as with the systematic velocity $\gamma$, that is, the RV of the barycenter of the binary system. The semi-amplitude $K$ depends on the orbital period $P$, and the minimum semi-major axis $a\sin(i)$ of the Keplerian orbit of the star, which is typically inclined to the plane of the sky by the inclination angle $i$:
\begin{equation}
K=\dfrac{2\pi\cdot a\sin(i)}{P\cdot \sqrt{1-e^{2}}}.
\end{equation}
We determined the orbital elements and their uncertainties for all observed spectroscopic binaries by fitting spectroscopic orbital solutions on the obtained RV data of the targets, using the method of least squares. Therefore, we calculated and minimized the chi-squared value and started the fitting procedure from an initial orbital period, derived by three different period determination algorithms, namely the normalized periodogram (\cite{lomb1976} \& \cite{scargle1982}), the string length (\cite{dworetsky1983}), and the phase-dispersion-minimization method (\cite{marraco1980}) with the spectroscopic binary solver (\cite{johnson2004}). All these methods yield very similar results, whose average was adopted as initial period for the orbit determination. \\
For CO\,Cam $(e=0.000\pm0.012)$, HR\,791 $(e=0.000\pm0.012)$, HR\,1401 $(e=0.003\pm0.008)$ and 71\,Dra $(e=0.000\pm0.008)$, the obtained orbital solutions are consistent with circular orbits. Therefore, in these cases the orbit fitting was repeated using a fixed orbital eccentricity $e=0$ and argument of periastron $\omega=0^{\circ}$.
The derived spectroscopic orbital elements of the systems with their uncertainties are summarized in Tab.\,\ref{circular} for the circular systems and Tab.\,\ref{elements} for the eccentric systems, respectively. The fitted and phase-folded RV curves of the determined orbits are illustrated, together with the RV data of all observed spectroscopic binaries, in Fig.\,\ref{fig:plot}.

\begin{table*}
	\caption{Derived orbital elements of the spectroscopic binaries with circular orbits ($e=0$, $\omega=0^{\circ}$).}
	\centering
	\begin{tabular}{ccccc}
		\hline
		Target          & $P$                   & $T$                      & $K$              & $\gamma$                         \\
		& $[\text{d}]$      				  & $[\text{BJD}-2450000]$ & $[$km/s$]$     & $[$km/s$]$                    \\
		\hline
		CO\,Cam			& $~~~1.271\pm0.001$  & $7430.93\pm0.01$       & $72.24\pm0.31$ & $\,\,\,+0.66\pm0.21$  \\
		HR\,791	 		& $~~~2.537\pm0.001$  & $7430.27\pm0.01$       & $55.40\pm0.44$ & $\,\,\,+3.93\pm0.32$  \\
		HR\,1401 		& $~~~4.194\pm0.001$  & $7430.96\pm0.01$       & $32.76\pm0.24$ & $+13.15\pm0.17$       \\
		71\,Dra	 		& $~~~5.298\pm0.001$  & $7428.01\pm0.02$       & $48.33\pm0.27$ & $\,\,\,-6.84\pm0.21$ \\
		\hline \\
	\end{tabular}                                              		
	\label{circular}
\end{table*}

\begin{table*}
\caption{Derived orbital elements of the spectroscopic binaries with eccentric orbits.}
\centering
\begin{tabular}{ccccccc}
		\hline
		Target          & $P$                   & $T$                      & $K$              & $\gamma$             & $\omega$        & $e$               \\
		& $[\text{d}]$        & $[\text{BJD}-2450000]$ & $[$km/s$]$     & $[$km/s$]$           & $[^{\circ}]$    &                 \\
		\hline
		$\alpha$\,Dra   & $~\,51.440\pm0.024$ & $7406.38\pm0.09$       & $47.48\pm0.21$ & $\,-13.50\pm0.12$    & $~\,21.8\pm0.6$ & $0.426\pm0.004$ \\
		$\omega$\,Cas	& $~\,69.623\pm0.058$ & $7428.18\pm0.22$       & $29.68\pm0.22$ & $-23.51\pm0.12$      & $~\,29.3\pm1.2$ & $0.305\pm0.006$ \\
		OS\,UMa			& $~\,89.036\pm0.115$ & $7359.05\pm0.54$       & $23.83\pm0.19$ & $\,\,\,-2.47\pm0.11$ & $183.1\pm2.3$   & $0.161\pm0.008$ \\
		$\varphi$\,Dra  & $128.165\pm0.056$   & $6956.94\pm0.54$       & $27.29\pm0.31$ & $-16.61\pm0.11$      & $134.8\pm0.9$   & $0.675\pm0.004$ \\
		\hline \\
	\end{tabular}                                              		
	\label{elements}
\end{table*}

\section{Discussion}	

In the course of our RV monitoring campaign of selected spectroscopic binary systems, a large number of RV measurements (more than 40, spanning 2 up to 179 orbital periods) have been collected with FLECHAS for all observed eight targets sufficiently. For these systems with nondefinite orbital solutions (all with orbit grades of less than 5, listed in the SB9) we were able to redetermine and/or to constrain their spectroscopic orbital elements, which are summarized in Tab.\,\ref{circular} and Tab.\,\ref{elements}.

For all systems, orbital solutions were determined with the least-squares fitting of spectroscopic orbits on the given RV measurements, which all exhibit reduced chi-squared values of $\chi^{2}_{red}\sim1$. This yields phase-folded RV curves, which are homogeneously covered with data points, whose uncertainties are all significantly smaller (by factors of 25 to 50) compared to the derived semi-amplitudes of the observed spectroscopic binary systems. Hence, for all systems the RV variation of their primary stars, induced by the motion of the stars around the barycenters of the systems, could be well detected and modeled by Keplerian motion.

For all but one system, the derived orbital periods deviate from the values given in the SB9 only on the per mille level. The largest deviation of 5 per mille was found for $\omega$\,Cas (a $\text{Grade}=3$ system). In the case of $\varphi$\,Dra, which is classified as a $\text{Grade}=1$ system in the SB9 catalog, we obtain an orbital period of $128.165\pm0.025$\,days, which is about $\sim4.8$ times longer than the given period in the SB9 catalog. Hence, the orbital elements of this spectroscopic binary system need significant modification to fit our RV data. Actually, besides the official value of 26.768\,days in the SB9 for $\varphi$\,Dra, a longer period of 127.85\,d is noted (uncertainty is not given), which is in good agreement with the orbital period of the system derived from our RV measurements. Furthermore, our orbit solution of the $\varphi$\,Dra system also exhibits a significantly higher eccentricity compared to the orbit given in the catalog.

In contrast of $\varphi$\,Dra, the obtained orbital eccentricities of all the other observed spectroscopic binary systems are consistent with those given in the SB9 catalog, with deviations of less than 0.04. All observed spectroscopic binary systems with orbital periods of only a few days exhibit almost circular orbits, while the orbits of systems with periods of a few dozens of days are significantly more eccentric (\cite{mayor1984}), up to $e=0.675$, in the case of $\varphi$\,Dra.

Also, the determined semi-amplitudes of the binary systems are consistent on the 1\,km/s level with the values listed in the SB9, except for the $\text{Grade}=3$ system CO\,Cam, whose semi-amplitude is 9\,km/s larger in our orbital solution compared to the value published in the SB9. The same result was found for the systematic velocities of all observed spectroscopic binaries, except for the $\text{Grade}=3$ system HR\,1401, whose systematic velocity derived by us deviates by 4\,km/s from the orbital solution published in the SB9.

As expected, the orbital solutions of all $\text{Grade}\geq2$ systems agree well with the spectroscopic orbits obtained from our RV measurements, and only exhibit smaller deviations. Our RV measurements constrain the orbital solutions for these spectroscopic binaries, which therefore should be classified as $\text{Grade}=5$ systems from now on. In contrast, the orbital elements of the $\text{Grade}=1$ system $\varphi$\,Dra, listed in the SB9, were all proven to be incorrect, and our RV measurements could be used to redetermine the spectroscopic orbital solution of this system.

From the determined orbital elements of the observed spectroscopic binary systems, we can derive their mass function $f(m)$  and the minimum semi-major axis $a\sin(i)$ of the orbits of their primary components, which are summarized in Tab.\,\ref{fma}.

\begin{table}[h!]
\caption{Derived mass functions $f(m)$ and minimum semi-major axes $a\sin(i)$ of all observed spectroscopic binaries.}
	\centering
	\begin{tabular}{ccc}
		\hline
		Target          & $f(m)$               & $a\sin(i)$         \\
		& $[\text{M}_{\odot}]$ & $[\text{au}]$     \\
		\hline
		CO\,Cam			& $0.050\pm0.001$      & $0.0084\pm0.0001$ \\
		HR\,791	 		& $0.045\pm0.002$      & $0.0129\pm0.0002$ \\
		HR\,1401 		& $0.015\pm0.001$      & $0.0126\pm0.0001$ \\
		71\,Dra	 		& $0.062\pm0.002$      & $0.0235\pm0.0002$ \\
		$\alpha$\,Dra   & $0.422\pm0.008$      & $0.2031\pm0.0011$ \\
		$\omega$\,Cas	& $0.163\pm0.005$      & $0.1809\pm0.0017$ \\
		OS\,UMa			& $0.120\pm0.004$      & $0.1925\pm0.0021$ \\
		$\varphi$\,Dra  & $0.108\pm0.005$      & $0.2371\pm0.0030$ \\
		\hline \\
	\end{tabular}                                              		
	\label{fma}
\end{table}

The mass function of a binary system and the minimum semi-major axis of its primary star are both dependent on the orbital eccentricity $e$, period $P$, and semi-amplitude $K$ given by
\begin{equation}
f(m)= \dfrac{M_{2}^{3}\cdot\sin^{3}(i)}{(M_{1}+M_{2})^{2}} = \dfrac{(1-e^{2})^{3/2}\cdot P \cdot K^{3}}{2\pi G},
\end{equation}
\begin{equation}
a\sin(i) = \dfrac{K\cdot P\cdot \sqrt{1-e^{2}}}{2\pi}.
\end{equation}
The observed spectroscopic binary systems exhibit mass-functions in the range between 0.015\,$\text{M}_{\odot}$ and 0.422\,$\text{M}_{\odot}$ with minimum semi-major axes of their primaries ranging between 0.0084\,au and 0.2371\,au. While the mass functions and minimum semi-major axes of HR\,791, HR\,1401, 71\,Dra and $\omega$\,Cas agree very well with the values listed in the SB9, those of $\varphi$\,Dra and CO\,Cam in particular deviate significantly. In the case of $\varphi$\,Dra, this deviation is due to the longer orbital period derived from our orbit fit, while for CO\,Cam the deviations result from a larger RV semi-amplitude. Furthermore, for $\alpha$\,Dra and OS\,UMa the mass functions and minimum semi-major axes deviate from the SB9 value by 17\,\% and 6\,\%, respectively, because of different RV semi-amplitudes.

As our RV monitoring campaign is an ongoing program, further RV measurements of spectroscopic binary systems will be undertaken in the future, and we want to collect RV measurements of all the remaining spectroscopic binaries from our original target list, which exhibit smaller RV semi-amplitudes, until precise orbital solutions for all these systems can be derived from our RV data. Furthermore, we plan  to extend the project and monitor additional circumpolar spectroscopic binaries with nondefinite orbital solutions, which are located at lower declinations. The new RV measurements and derived orbital solutions, which will be obtained and derived in the course of our spectroscopy campaign in the future, together with the data presented in this paper, will be available online at \texttt{VizieR} (\cite{ochsenbein2000}).

\begin{acknowledgement}
We want to thank all observers who have participated in some of the observations of this project, obtained at the University Observatory Jena, in particular S. Masda, H. Gilbert, S. Buder and B. Dincel. This publication makes use of data products of the \texttt{SIMBAD} and \texttt{VizieR} databases, operated at CDS, Strasbourg, France.
\end{acknowledgement}

\newpage

\appendix\section{Radial velocity measurements}

\begin{table}[h!] \caption{RV measurements of CO\,Cam.}
	\begin{center}\begin{tabular}{cc}
		\hline
		Date of Obs.         & RV	      \\
		$\text{BJD}-2450000$ & $[$km/s$]$ \\
		\hline
7431.38824	& $-46.3\pm1.4$\\
7485.35936	& $+33.3\pm1.4$\\
7486.36825	& $-52.0\pm1.4$\\
7486.46645	& $-24.1\pm1.4$\\
7499.62351	& $+70.2\pm1.4$\\
7500.32658	& $-57.3\pm1.4$\\
7500.59242	& $+28.0\pm1.4$\\
7507.33987	& $+55.6\pm1.4$\\
7507.59890	& $-28.9\pm1.4$\\
7518.40136	& $+29.1\pm1.4$\\
7518.57816	& $+69.4\pm1.4$\\
7520.42959	& $-60.1\pm1.4$\\
7520.55491	& $-73.5\pm1.4$\\
7527.38213	& $+53.5\pm1.4$\\
7528.49391	& $+05.8\pm1.4$\\
7535.36723	& $+39.1\pm1.4$\\
7549.46284	& $-04.0\pm1.4$\\
7549.52179	& $-22.4\pm1.4$\\
7555.46304	& $+70.6\pm1.4$\\
7562.40211	& $-65.5\pm1.4$\\
7562.52331	& $-70.8\pm1.4$\\
7563.40229	& $+13.2\pm1.4$\\
7563.54301	& $-34.3\pm1.4$\\
7575.54729	& $+16.1\pm1.4$\\
7576.48260	& $-72.8\pm1.4$\\
7588.37243	& $+51.5\pm1.4$\\
7588.52811	& $+71.7\pm1.4$\\
7589.37415	& $-39.2\pm1.4$\\
7589.52940	& $+15.3\pm1.4$\\
7590.39096	& $-69.5\pm1.4$\\
7591.39757	& $+03.8\pm1.4$\\
7591.55009	& $-49.3\pm1.4$\\
7596.41929	& $+22.6\pm1.4$\\
7597.39022	& $+70.7\pm1.4$\\
7611.47725	& $+68.9\pm1.4$\\
7611.60414	& $+42.8\pm1.4$\\
7618.33899	& $-70.5\pm1.4$\\
7618.59800	& $-42.3\pm1.4$\\
7619.36532	& $-01.8\pm1.4$\\
7619.60392	& $-67.0\pm1.4$\\
7638.31841	& $+38.1\pm1.4$\\
7638.51053	& $-28.2\pm1.4$\\
7644.46335	& $+70.3\pm1.4$\\
7646.62030	& $-22.3\pm1.4$\\
7659.28240	& $-39.0\pm1.4$\\
		\hline \\
	\end{tabular}\end{center}                                              		
	\label{rv_cocam}
\end{table}

\begin{table}[t!] \caption{RV measurements of HR\,791.}
	\begin{center}\begin{tabular}{cc}
		\hline
		Date of Obs.         & RV	      \\
		$\text{BJD}-2450000$ & $[$km/s$]$ \\
		\hline
7431.35565	& $	-45.4\pm2.1$\\
7431.48233	& $	-54.4\pm2.1$\\
7485.33239	& $	-07.8\pm2.1$\\
7486.34242	& $	+47.3\pm2.1$\\
7486.44723	& $	+39.9\pm2.1$\\
7499.59356	& $	-18.4\pm2.1$\\
7500.34968	& $	-34.9\pm2.1$\\
7500.57077	& $	-07.4\pm2.1$\\
7507.38750	& $	-40.2\pm2.1$\\
7507.57836	& $	-52.5\pm2.1$\\
7518.37140	& $	+00.8\pm2.1$\\
7518.55775	& $	+21.6\pm2.1$\\
7520.40708	& $	-49.4\pm2.1$\\
7520.53198	& $	-45.2\pm2.1$\\
7527.42254	& $	-10.6\pm2.1$\\
7528.51179	& $	+00.4\pm2.1$\\
7535.41911	& $	-50.0\pm2.1$\\
7549.44273	& $	+52.3\pm2.1$\\
7549.53912	& $	+61.3\pm2.1$\\
7555.44365	& $	-28.0\pm2.1$\\
7562.44093	& $	+49.9\pm2.1$\\
7562.50411	& $	+40.6\pm2.1$\\
7563.43937	& $	-52.0\pm2.1$\\
7563.52551	& $	-50.2\pm2.1$\\
7570.51037	& $	-03.9\pm2.1$\\
7575.52838	& $	-02.1\pm2.1$\\
7576.45255	& $	-35.7\pm2.1$\\
7576.54044	& $	-24.6\pm2.1$\\
7588.40847	& $	-26.2\pm2.1$\\
7588.54441	& $	-40.4\pm2.1$\\
7589.39046	& $	-00.6\pm2.1$\\
7589.54588	& $	+18.0\pm2.1$\\
7590.41140	& $	+39.3\pm2.1$\\
7591.44834	& $	-51.3\pm2.1$\\
7591.56621	& $	-44.0\pm2.1$\\
7596.43751	& $	-51.9\pm2.1$\\
7597.42024	& $	+49.5\pm2.1$\\
7611.45042	& $	-44.1\pm2.1$\\
7611.59040	& $	-53.2\pm2.1$\\
7616.36420	& $	-33.5\pm2.1$\\
7617.37327	& $	+06.4\pm2.1$\\
7617.58428	& $	+32.4\pm2.1$\\
7618.35774	& $	+38.0\pm2.1$\\
7618.61459	& $	+02.7\pm2.1$\\
7619.35749	& $	-49.9\pm2.1$\\
7619.59732	& $	-31.2\pm2.1$\\
7638.31135	& $	+57.6\pm2.1$\\
7638.53898	& $	+46.7\pm2.1$\\
7644.31020	& $	-37.2\pm2.1$\\
7644.43418	& $	-46.7\pm2.1$\\
7646.48445	& $	+04.0\pm2.1$\\
7659.35204	& $	-15.8\pm2.1$\\
		\hline \\
	\end{tabular}\end{center}                                              		
	\label{rv_hr971}
\end{table}

\begin{table}[t!] \caption{RV measurements of HR\,1401.}
	\begin{center}\begin{tabular}{cc}
		\hline
		Date of Obs.         & RV	      \\
		$\text{BJD}-2450000$ & $[$km/s$]$ \\
		\hline
7431.36413	& $	+40.0\pm1.2$\\
7431.49177	& $	+34.5\pm1.2$\\
7485.34862	& $	+44.8\pm1.2$\\
7486.35091	& $	+21.7\pm1.2$\\
7486.45609	& $	+16.9\pm1.2$\\
7499.60181	& $	-08.8\pm1.2$\\
7500.35764	& $	-18.7\pm1.2$\\
7500.58000	& $	-13.4\pm1.2$\\
7507.39571	& $	+17.8\pm1.2$\\
7507.58639	& $	+09.3\pm1.2$\\
7518.38131	& $	+32.7\pm1.2$\\
7518.56410	& $	+35.5\pm1.2$\\
7520.41798	& $	-03.5\pm1.2$\\
7520.54136	& $	-07.6\pm1.2$\\
7527.49823	& $	+45.3\pm1.2$\\
7528.51890	& $	+12.7\pm1.2$\\
7535.43266	& $	+42.7\pm1.2$\\
7549.45070	& $	+14.4\pm1.2$\\
7549.53148	& $	+11.3\pm1.2$\\
7555.45150	& $	-00.7\pm1.2$\\
7562.44834	& $	-04.6\pm1.2$\\
7562.51140	& $	-09.4\pm1.2$\\
7563.44625	& $	-14.4\pm1.2$\\
7563.53199	& $	-12.8\pm1.2$\\
7570.51846	& $	+08.2\pm1.2$\\
7575.53568	& $	-18.8\pm1.2$\\
7576.46055	& $	+02.4\pm1.2$\\
7576.54788	& $	+03.3\pm1.2$\\
7588.41480	& $	-17.8\pm1.2$\\
7588.55076	& $	-15.5\pm1.2$\\
7589.39766	& $	+19.0\pm1.2$\\
7589.55216	& $	+24.6\pm1.2$\\
7590.41817	& $	+47.4\pm1.2$\\
7591.45481	& $	+09.3\pm1.2$\\
7591.57239	& $	+01.7\pm1.2$\\
7596.44108	& $	-19.7\pm1.2$\\
7597.41268	& $	+00.9\pm1.2$\\
7611.45663	& $	+43.6\pm1.2$\\
7611.59654	& $	+42.7\pm1.2$\\
7616.35430	& $	+23.0\pm1.2$\\
7618.34946	& $	-01.4\pm1.2$\\
7618.60677	& $	+10.5\pm1.2$\\
7619.34775	& $	+41.7\pm1.2$\\
7619.58832	& $	+45.7\pm1.2$\\
7638.30298	& $	-16.9\pm1.2$\\
7638.53079	& $	-20.3\pm1.2$\\
7644.30046	& $	+33.7\pm1.2$\\
7644.44064	& $	+41.1\pm1.2$\\
7646.50562	& $	-11.4\pm1.2$\\
7659.33373	& $	-17.1\pm1.2$\\
		\hline \\
\\
\\
	\end{tabular}\end{center}                                              		
	\label{rv_hr1401}
\end{table}

\begin{table}[t!] \caption{RV measurements of 71\,Dra.}
	\begin{center}\begin{tabular}{cc}
		\hline
		Date of Obs.         & RV	      \\
		$\text{BJD}-2450000$ & $[$km/s$]$ \\
		\hline
7431.41221	& $	-39.1\pm1.4$\\
7485.37684	& $	+14.1\pm1.4$\\
7486.39515	& $	+41.9\pm1.4$\\
7486.51331	& $	+39.2\pm1.4$\\
7499.57297	& $	-52.1\pm1.4$\\
7500.34119	& $	-34.6\pm1.4$\\
7500.55710	& $	-24.2\pm1.4$\\
7507.36448	& $	+39.8\pm1.4$\\
7507.57273	& $	+43.6\pm1.4$\\
7518.41921	& $	+39.0\pm1.4$\\
7518.58686	& $	+32.1\pm1.4$\\
7520.43935	& $	-53.2\pm1.4$\\
7520.56459	& $	-54.1\pm1.4$\\
7527.40011	& $	-02.8\pm1.4$\\
7528.50668	& $	+43.0\pm1.4$\\
7535.39262	& $	-14.3\pm1.4$\\
7549.47435	& $	+38.1\pm1.4$\\
7549.51487	& $	+38.1\pm1.4$\\
7555.47357	& $	+39.0\pm1.4$\\
7562.42136	& $	-40.8\pm1.4$\\
7562.49907	& $	-41.7\pm1.4$\\
7563.42083	& $	-51.1\pm1.4$\\
7563.52062	& $	-47.6\pm1.4$\\
7570.47091	& $	+31.2\pm1.4$\\
7575.55873	& $	+22.9\pm1.4$\\
7576.50313	& $	+40.3\pm1.4$\\
7588.38906	& $	-15.8\pm1.4$\\
7588.53940	& $	-21.1\pm1.4$\\
7589.38556	& $	-54.4\pm1.4$\\
7589.54103	& $	-53.8\pm1.4$\\
7590.38080	& $	-37.7\pm1.4$\\
7591.40884	& $	+18.8\pm1.4$\\
7591.56145	& $	+27.3\pm1.4$\\
7596.43087	& $	+06.2\pm1.4$\\
7597.40165	& $	+40.0\pm1.4$\\
7602.44154	& $	+35.6\pm1.4$\\
7611.43121	& $	-42.0\pm1.4$\\
7611.61722	& $	-33.2\pm1.4$\\
7616.37540	& $	-50.7\pm1.4$\\
7617.34675	& $	-11.4\pm1.4$\\
7617.59434	& $	+02.3\pm1.4$\\
7618.33441	& $	+35.6\pm1.4$\\
7618.59222	& $	+40.2\pm1.4$\\
7619.33842	& $	+29.1\pm1.4$\\
7619.58303	& $	+15.9\pm1.4$\\
7638.29618	& $	-23.0\pm1.4$\\
7638.48889	& $	-15.4\pm1.4$\\
7644.29289	& $	+14.4\pm1.4$\\
7644.48369	& $	+25.3\pm1.4$\\
7646.44431	& $	+00.4\pm1.4$\\
7659.26740	& $	-37.0\pm1.4$\\
		\hline \\
	\end{tabular}\end{center}                                              		
	\label{rv_71dra}
\end{table}

\begin{table}[t!] \caption{RV measurements of $\alpha$\,Dra.}
	\begin{center}\begin{tabular}{cc}
		\hline
		Date of Obs.         & RV	      \\
		$\text{BJD}-2450000$ & $[$km/s$]$ \\
		\hline
7431.39620	& $	-39.8\pm0.8$\\
7456.51247	& $	+51.8\pm0.8$\\
7457.50988	& $	+50.9\pm0.8$\\
7464.42365	& $	-17.0\pm0.8$\\
7465.47306	& $	-21.7\pm0.8$\\
7485.36787	& $	-37.4\pm0.8$\\
7486.37752	& $	-36.8\pm0.8$\\
7491.33144	& $	-27.6\pm0.8$\\
7499.63239	& $	+00.3\pm0.8$\\
7500.30910	& $	+04.0\pm0.8$\\
7507.34876	& $	+50.5\pm0.8$\\
7511.35961	& $	+28.2\pm0.8$\\
7518.35851	& $	-28.4\pm0.8$\\
7519.50551	& $	-31.5\pm0.8$\\
7520.36909	& $	-34.6\pm0.8$\\
7527.39125	& $	-41.7\pm0.8$\\
7528.54961	& $	-41.9\pm0.8$\\
7535.37655	& $	-37.6\pm0.8$\\
7546.48331	& $	-17.2\pm0.8$\\
7547.46095	& $	-13.1\pm0.8$\\
7549.42603	& $	-05.0\pm0.8$\\
7555.42805	& $	+30.0\pm0.8$\\
7562.41092	& $	+35.0\pm0.8$\\
7563.41089	& $	+20.8\pm0.8$\\
7568.39729	& $	-23.3\pm0.8$\\
7575.50080	& $	-40.4\pm0.8$\\
7576.49245	& $	-40.8\pm0.8$\\
7588.37844	& $	-35.7\pm0.8$\\
7589.49422	& $	-35.7\pm0.8$\\
7590.32846	& $	-33.6\pm0.8$\\
7591.38119	& $	-32.1\pm0.8$\\
7597.38293	& $	-19.4\pm0.8$\\
7602.34230	& $	-00.4\pm0.8$\\
7611.49387	& $	+52.9\pm0.8$\\
7616.39098	& $	+03.6\pm0.8$\\
7617.45327	& $	-05.1\pm0.8$\\
7618.38174	& $	-13.4\pm0.8$\\
7619.37289	& $	-19.7\pm0.8$\\
7638.50127	& $	-38.4\pm0.8$\\
7644.47055	& $	-30.0\pm0.8$\\
7659.27274	& $	+37.3\pm0.8$\\
		\hline \\
\\
\\
\\
\\
\\
\\
\\
\\
\\
\\
	\end{tabular}\end{center}                                              		
	\label{rv_alfdra}
\end{table}

\begin{table}[t!] \caption{RV measurements of $\omega$\,Cas.}
	\begin{center}\begin{tabular}{cc}
		\hline
		Date of Obs.         & RV	      \\
		$\text{BJD}-2450000$ & $[$km/s$]$ \\
		\hline
7431.04008	& $	+00.2\pm0.8$\\
7442.25414	& $	-39.1\pm0.8$\\
7456.33374	& $	-45.3\pm0.8$\\
7457.30378	& $	-45.4\pm0.8$\\
7464.43439	& $	-41.0\pm0.8$\\
7465.27342	& $	-39.3\pm0.8$\\
7485.33989	& $	-05.7\pm0.8$\\
7486.33541	& $	-02.9\pm0.8$\\
7491.33758	& $	+11.1\pm0.8$\\
7499.58638	& $	+05.4\pm0.8$\\
7500.56372	& $	+00.4\pm0.8$\\
7507.37969	& $	-26.9\pm0.8$\\
7511.33749	& $	-36.6\pm0.8$\\
7518.34224	& $	-44.4\pm0.8$\\
7519.48114	& $	-44.7\pm0.8$\\
7520.39168	& $	-44.6\pm0.8$\\
7527.41194	& $	-43.8\pm0.8$\\
7528.56423	& $	-43.3\pm0.8$\\
7535.40529	& $	-39.3\pm0.8$\\
7546.45869	& $	-24.7\pm0.8$\\
7547.43634	& $	-22.1\pm0.8$\\
7549.40154	& $	-19.7\pm0.8$\\
7555.40609	& $	-06.0\pm0.8$\\
7562.43427	& $	+11.9\pm0.8$\\
7563.43270	& $	+13.9\pm0.8$\\
7575.46018	& $	-20.6\pm0.8$\\
7576.44445	& $	-24.0\pm0.8$\\
7588.40185	& $	-44.1\pm0.8$\\
7589.51029	& $	-45.4\pm0.8$\\
7590.40524	& $	-46.0\pm0.8$\\
7591.44208	& $	-44.4\pm0.8$\\
7597.42556	& $	-43.5\pm0.8$\\
7597.57628	& $	-43.3\pm0.8$\\
7602.42013	& $	-41.7\pm0.8$\\
7611.44409	& $	-32.6\pm0.8$\\
7616.38317	& $	-23.5\pm0.8$\\
7617.44494	& $	-22.7\pm0.8$\\
7618.36495	& $	-19.1\pm0.8$\\
7619.39356	& $	-17.7\pm0.8$\\
7638.54635	& $	+04.3\pm0.8$\\
7644.44410 	& $	-20.1\pm0.8$\\
7646.47627	& $	-25.1\pm0.8$\\
7659.34373	& $	-45.3\pm0.8$\\	
		\hline \\
	\end{tabular}\end{center}                                              		
	\label{rv_omecas}
\end{table}

\begin{table}[t!] \caption{RV measurements of OS\,UMa.}
	\begin{center}\begin{tabular}{cc}
		\hline
		Date of Obs.         & RV	      \\
		$\text{BJD}-2450000$ & $[$km/s$]$ \\
		\hline
7431.38140	& $	-09.4\pm0.7$\\
7442.26695	& $	-28.1\pm0.7$\\
7456.27464	& $	-21.8\pm0.7$\\
7457.31735	& $	-19.8\pm0.7$\\
7464.44664	& $	-08.4\pm0.7$\\
7465.46437	& $	-06.1\pm0.7$\\
7485.32577	& $	+15.9\pm0.7$\\
7486.36029	& $	+16.1\pm0.7$\\
7491.32162	& $	+17.7\pm0.7$\\
7499.61433	& $	+14.8\pm0.7$\\
7500.31617	& $	+15.5\pm0.7$\\
7507.40505	& $	+09.3\pm0.7$\\
7511.34835	& $	+04.3\pm0.7$\\
7518.39152	& $	-05.5\pm0.7$\\
7519.49408	& $	-06.8\pm0.7$\\
7520.37536	& $	-09.9\pm0.7$\\
7527.46506	& $	-21.1\pm0.7$\\
7528.55471	& $	-24.3\pm0.7$\\
7535.45215	& $	-29.1\pm0.7$\\
7546.47118	& $	-20.1\pm0.7$\\
7547.44870	& $	-18.6\pm0.7$\\
7549.41409	& $	-15.7\pm0.7$\\
7555.41660	& $	-03.8\pm0.7$\\
7562.45912	& $	+06.7\pm0.7$\\
7563.46016	& $	+08.7\pm0.7$\\
7575.46980	& $	+17.9\pm0.7$\\
7575.51646	& $	+17.6\pm0.7$\\
7576.47049	& $	+16.2\pm0.7$\\
7588.42613	& $	+15.6\pm0.7$\\
7589.51650	& $	+14.3\pm0.7$\\
7590.42777	& $	+15.1\pm0.7$\\
7591.46461	& $	+12.7\pm0.7$\\
7597.50344	& $	+08.8\pm0.7$\\
7611.46424	& $	-12.1\pm0.7$\\
7618.37188	& $	-25.1\pm0.7$\\
7619.38024	& $	-26.1\pm0.7$\\
7638.52103	& $	-14.1\pm0.7$\\
7644.44990	& $	-04.0\pm0.7$\\
7646.61182	& $	-01.0\pm0.7$\\
7659.32290	& $	+14.1\pm0.7$\\	
		\hline \\
\\
\\
\\
	\end{tabular}\end{center}                                              		
	\label{rv_osuma}
\end{table}

\begin{table*}[h!]
\caption{RV measurements of $\varphi$\,Dra.}
 	\begin{center}\begin{tabular}{cccc}
		\hline
		Date of Obs.         & RV	      & Date of Obs.         & RV         \\
		$\text{BJD}-2450000$ & $[$km/s$]$ & $\text{BJD}-2450000$ & $[$km/s$]$ \\
		\hline
 7076.60709	& $	-07.7\pm0.9$ & 7289.53045	& $	-07.8\pm0.9$\\
 7079.63092	& $	-14.7\pm0.9$ & 7290.27209	& $	-05.8\pm0.9$\\
 7080.53165	& $	-17.8\pm0.9$ & 7293.39512	& $	-07.3\pm0.9$\\
 7082.40602	& $	-27.9\pm0.9$ & 7294.28817	& $	-06.4\pm0.9$\\
 7100.49293	& $	-34.5\pm0.9$ & 7295.30253	& $	-06.6\pm0.9$\\
 7102.48464	& $	-32.5\pm0.9$ & 7296.39483	& $	-06.2\pm0.9$\\
 7106.44076	& $	-28.3\pm0.9$ & 7297.26476	& $	-05.3\pm0.9$\\
 7122.43323	& $	-18.8\pm0.9$ & 7298.26216	& $	-05.9\pm0.9$\\
 7131.34166	& $	-16.1\pm0.9$ & 7299.39953	& $	-04.2\pm0.9$\\
 7135.50151	& $	-13.1\pm0.9$ & 7307.25747	& $	-04.5\pm0.9$\\
 7136.33505	& $	-13.0\pm0.9$ & 7359.51396	& $	-31.0\pm0.9$\\
 7149.58504	& $	-09.8\pm0.9$ & 7360.37400	& $	-29.9\pm0.9$\\
 7150.36169	& $	-11.9\pm0.9$ & 7364.27198	& $	-26.6\pm0.9$\\
 7158.58157	& $	-08.6\pm0.9$ & 7365.23393	& $	-26.8\pm0.9$\\
 7164.37657	& $	-08.5\pm0.9$ & 7366.33803	& $	-25.1\pm0.9$\\
 7178.38443	& $	-05.1\pm0.9$ & 7367.20762	& $	-24.7\pm0.9$\\
 7179.39581	& $	-03.2\pm0.9$ & 7396.32200	& $	-13.9\pm0.9$\\
 7184.35710	& $	-02.7\pm0.9$ & 7406.20814	& $	-11.1\pm0.9$\\
 7187.48292	& $	-01.9\pm0.9$ & 7409.36344	& $	-09.8\pm0.9$\\
 7189.41496	& $	-02.8\pm0.9$ & 7410.33114	& $	-08.5\pm0.9$\\
 7190.47897	& $	-03.2\pm0.9$ & 7422.29646	& $	-06.5\pm0.9$\\
 7198.45040	& $	-03.0\pm0.9$ & 7425.29230	& $	-06.4\pm0.9$\\
 7205.37426	& $	-07.5\pm0.9$ & 7431.40210	& $	-03.8\pm0.9$\\
 7206.38820	& $	-08.9\pm0.9$ & 7486.38454	& $	-33.2\pm0.9$\\
 7207.40064	& $	-13.0\pm0.9$ & 7499.56590	& $	-22.0\pm0.9$\\
 7210.39333	& $	-26.9\pm0.9$ & 7500.55010	& $	-22.6\pm0.9$\\
 7213.39737	& $	-49.9\pm0.9$ & 7507.35756	& $	-18.0\pm0.9$\\
 7214.39610	& $	-54.1\pm0.9$ & 7527.51212	& $	-12.6\pm0.9$\\
 7220.40472	& $	-48.4\pm0.9$ & 7528.57514	& $	-11.5\pm0.9$\\
 7221.44518	& $	-46.7\pm0.9$ & 7535.38500	& $	-09.7\pm0.9$\\
 7233.39523	& $	-28.6\pm0.9$ & 7563.42001	& $	-04.1\pm0.9$\\	
		\hline \\
	\end{tabular}\end{center}                                              		
	\label{rv_phidra}
\end{table*}

\end{document}